\documentstyle[12pt,aaspp4,mathptm,epsfig,rotating]{article}

\begin{document}

\title{The structure of protostellar envelopes derived from\\
submillimeter continuum images}
\author{Claire J. Chandler}
\and
\author{John S. Richer}
\affil{Department of Physics, University of Cambridge, Cavendish
Laboratory, Madingley Road, Cambridge CB3 0HE, United Kingdom}
\affil{E-mail: cjc@mrao.cam.ac.uk, jsr@mrao.cam.ac.uk}

\begin{abstract}

High dynamic range imaging of submillimeter dust emission from the
envelopes of eight young protostars in the Taurus and Perseus
star-forming regions has been carried out using the SCUBA
submillimeter camera on the James Clerk Maxwell Telescope.  Good
correspondence between the spectral classifications of the
protostars and the spatial distributions of their dust emission is
observed, in the sense that those with cooler spectral energy
distributions also have a larger fraction of the submillimeter flux
originating in an extended envelope compared with a disk.  This
results from the cool sources having more massive envelopes rather
than warm sources having larger disks.  Azimuthally-averaged radial
profiles of the dust emission are used to derive the power-law index
of the envelope density distributions, $p$ (defined by $\rho \propto
r^{-p}$), and most of the sources are found to have values of $p$
consistent with those predicted by models of cloud collapse.
However, the youngest protostars in our sample, L1527 and HH211-mm,
deviate significantly from the theoretical predictions, exhibiting
values of $p$ somewhat lower than can be accounted for by existing
models.  For L1527 heating of the envelope by shocks where the
outflow impinges on the surrounding medium may explain our result. 
For HH211-mm another explanation is needed, and one possibility is
that a shallow density profile is being maintained in the outer
envelope by magnetic fields and/or turbulence.  If this is the case
star formation must be determined by the rate at which the support
is lost from the cloud, rather than the hydrodynamical properties of
the envelope, such as the sound speed.

\end{abstract}


\section{Introduction}
\label{introduction}

One of the fundamental problems associated with assembling an
observationally-based framework for early protostellar evolution is
that the photospheres of deeply-embedded protostars are highly
obscured, and the techniques used for establishing the properties of
optically-visible young stellar objects, such as comparison with
model isochrones in the Hertzsprung-Russell diagram, are not
available.  Instead, we are limited to measuring only (1) the
interaction of the protostar with the surrounding molecular cloud,
and (2) the luminosity of the protostar/disk system embedded at the
center.  Neither of these can provide a value for the mass, radius,
or age of the protostar directly, without making several assumptions
of how the central mass is being accumulated as a function of time. 
For example, one possible measure of the age of a protostar may be
the dynamical age of its outflow, but it is hard to tie this to the
actual age of the protostar without knowing exactly when the mass
loss began, or exactly how the outflow propagates through the
interstellar medium.  Similarly, the bolometric luminosity, $L_{\rm
bol}$, has been used to infer protostellar ages (Barsony 1994), by
assuming it derives entirely from accretion so that $L_{\rm bol} =
GM_*(t)\dot{M}/R_*$, where $\dot{M}$ is the accretion rate, the
stellar mass $M_*(t) = \int\dot{M}dt$, and $R_*$ is the (assumed)
stellar radius.  The protostellar age is then approximately
$M_*(t)/\dot{M} = L_{\rm bol}R_*/G\dot{M}^2$, which will clearly
depend upon how $\dot{M}$ evolves with time.

Predictions for $\dot{M}(t)$ during cloud collapse vary
dramatically, and depend upon the initial density distribution
within the cloud.  The model most commonly cited in the literature
is the self-similar solution for the collapse of a singular
isothermal sphere (SIS), which is based on the work of Shu (1977)
and subsequent refinements to include slow rotation (Terebey, Shu,
\& Cassen 1984).  In this model collapse begins in a sphere with a
density distribution $\rho \propto r^{-2}$ at all radii.  Such a
configuration is highly unstable, and if perturbed, a region of
free-fall collapse spreads outwards from the center at the local
sound speed, $a$.  The rate at which material arrives at the
hydrostatic protostar forming at the center of the collapse flow is
therefore constant, with a value $\dot{M} \sim a^3/G$, where $G$ is
the gravitational constant.  On the other hand, self-similar
solutions for isothermal collapse beginning from unstable
configurations with density distributions flatter than the SIS have
time-varying accretion rates which can peak at approximately fifty
times the Shu value, declining thereafter (Larson 1969; Penston
1969; Hunter 1977; Whitworth \& Summers 1985).  Numerical
hydrodynamical simulations show results similar to those described
by the analytic solutions (Foster \& Chevalier 1993).

The dependence of $\dot{M}$ on the initial conditions from which
collapse begins is clearly extremely important for the subsequent
evolution of $L_{\rm bol}$ and $M_*$, affecting the numbers of
protostars detectable in a given stage of evolution, and their
derived statistical lifetimes.  Unfortunately, independent measures
of $\dot{M}(t)$ are not available for deeply-embedded protostars,
but the models do suggest other means of discriminating between the
various possibilities for early protostellar evolution and cloud
collapse.  After a hydrostatic protostar has formed at the centre of
a cloud core, the density distribution within the surrounding cloud
also depends on the initial conditions of the collapse.  In the
extreme case of the collapse of a SIS, material is stationary and
the density distribution remains proportional to $r^{-2}$ outside
the expanding region of infall, while directly inside the expansion
wave it becomes $r^{-1}$, steepening again to $r^{-3/2}$ as free
fall is established (Shu 1977).  For collapse initiated from flatter
density distributions the outer parts of the cloud also have $\rho
\propto r^{-2}$, but are already moving inwards, and match smoothly
to an expanding region of free fall having $\rho \propto r^{-3/2}$
(Whitworth \& Summers 1985; Foster \& Chevalier 1993).  These models
therefore suggest two potential observational tests of the initial
conditions, those of measuring the radial density profiles of cloud
cores and protostellar envelopes, and of determining the velocity
distribution to establish whether the outer parts are static or
infalling.  This latter test is particularly difficult once a
protostar has already formed, because outflow and multiple velocity
components along the line of sight confuse the interpretation of
spectral line emission.  Perhaps more promising as an initial test
of the collapse models is the measurement of radial density
profiles, particularly taking advantage of the simple radiative
transfer of millimeter and submillimeter dust emission.  This
technique has been successfully used for establishing that the
density profiles of starless cores are somewhat flattened within a
radius of approximately 4000~AU, having $\rho \propto r^{-1.2}$
rather than the $r^{-2}$ of the SIS (Ward-Thompson et al.\ 1994;
Andr\'e, Ward-Thompson, \& Motte 1996).

Once a dense core has lost all means of support and has begun to
collapse, and a hydrostatic protostar has formed at the center, then
the density profile in the surrounding envelope should be between
$r^{-3/2}$ and $r^{-2}$ for initial conditions comprising
non-singular isothermal spheres in unstable hydrostatic equilibria. 
Indeed, in this picture, we would expect to detect an evolution from
$r^{-2}$ in the envelopes of the youngest (Class~0) protostars to
the more shallow $r^{-3/2}$ in older sources.  For cloud cores which
are initially sufficiently centrally condensed the numerical
simulations of Foster \& Chevalier (1993) suggest that at late times
the hydrodynamics of the collapse resembles the Shu solution.  In
this case, the evolution of the density profile might be more
complicated, starting out with $r^{-2}$, and moving through a
somewhat flatter ($\rho \propto r^{-1}$) phase until $r^{-3/2}$
dominates.  For the few objects which have been studied to date, the
results remain confusing.  Some sources do show the $r^{-3/2}$
consistent with free-fall collapse (e.g., B335: Chandler \& Sargent
1993), but others seem to exhibit extremely flat profiles,
$r^{-0.5}$ (e.g., VLA1623: Andr\'e, Ward-Thompson, \& Barsony
1993).  A recent analysis of the shape of far-infrared/millimeter
dust spectral energy distributions (SEDs) suggests that $r^{-1}$ may
be appropriate for the Class~I sources TMC1, TMC1A, and Haro~6-10 in
Taurus (Chandler, Barsony, \& Moore 1998), and few objects show
density profiles as steep as the $r^{-2}$ expected for early times.

The constraints on the density distributions obtained so far have
been limited by the poor dynamic range achievable using
single-element detectors on submillimeter telescopes.  This
situation has been dramatically changed recently by the development
of submillimeter array cameras such as the Submillimetre Common-User
Bolometer Array (SCUBA) on the James Clerk Maxwell Telescope (JCMT),
which finally allows a detailed investigation of the distribution of
dust emission from protostellar envelopes, to search for evidence of
evolutionary and environmental effects in envelope structure.  We
have used SCUBA to image the submillimeter dust emission from
circumstellar material around eight protostars, four in the Taurus
molecular cloud and four in Perseus, and we use the observed
intensity profiles to derive density distributions in their
envelopes.  Submillimeter dust emission is particularly suitable
because the envelope emission is optically thin, but a problem
arises when considering all possible origins of the dust emission. 
Even young, deeply embedded protostellar systems are surrounded by
circumstellar disks (e.g., Chandler 1998), and this compact
component must be separated from the envelope if the structure of
the envelope is to be analyzed.  This requirement limits somewhat
the number of sources for which envelope structure can be studied.

\section{Source sample}

Since considerable information about both the compact disk component
and the geometry of any outflow cavity is required for a full
analysis of the observed dust distribution, only well-studied
objects have been observed in this current work.  This does, of
course, produce a highly-biased sample, but is clearly a necessary
first step.  Eight protostars have been selected, four from the
Taurus molecular cloud ($D = 140$~pc: Elias 1978; Kenyon, Dobrzycka,
\& Hartmann 1994), and four from the Perseus star forming region. 
The distance to Perseus is uncertain, with estimates ranging from
220~pc (Cernis 1990) to 350~pc (Herbig \& Jones 1983) reported in
the literature.  A distance of 350~pc for all the Perseus sources is
assumed in the analysis below, but the effect of using a smaller
distance on the final results is also investigated.  The sources
cover a range of spectral class and luminosity
(Table~\ref{source-sample}), and their outflow directions, opening
angles, and inclination angles, are all known.  The Taurus and
Perseus clouds are forming stars in somewhat different modes; the
protostars in Taurus are fairly isolated, and are predominantly
low-mass, while those in Perseus are more clustered and more
luminous (e.g., Ladd, Lada, \& Myers 1993).  The average temperature
of the molecular gas in the Perseus cloud is also a few degrees
higher than in Taurus (Ladd, Myers, \& Goodman 1994), which
translates into a higher sound speed for the gas in Perseus compared
with Taurus.

\section{Observations}

The submillimeter camera SCUBA on the 15-m JCMT on Mauna Kea,
Hawaii, comprises two bolometer arrays, a long-wavelength array
containing 37 elements optimized for operation in the 850 and
750~$\mu$m atmospheric windows, the other a 91-element
short-wavelength array with filters at 450 and 350~$\mu$m. 
Combinations of 850/450~$\mu$m and 750/350~$\mu$m may be observed
simultaneously through the use of a dichroic beam splitter.  The
bolometers are located in a hexagonal pattern, spaced two full
beamwidths apart, so the secondary mirror is ``jiggled'' to produce
a fully-sampled image.  Further technical details of observing with
SCUBA on the JCMT are described by Holland et al.\ (1999).  All
eight sources listed in Table~\ref{source-sample} were observed at
all four wavelengths in 1998 January, using a chop throw of
120~arcsec and a chop frequency of 7.8~Hz.  The chop direction for
each source was chosen to avoid both the outflow and possible
envelope emission perpendicular to the outflow.  L1551~IRS5 and
TMC1A were also observed at 850~$\mu$m, with a chop throw 120~arcsec
in azimuth, in service mode in 1997 July and August.  The
atmospheric transmission was monitored by performing sky dips at
regular intervals, which, combined with measurements of the 225~GHz
opacity obtained from the tipping radiometer located at the Caltech
Submillimeter Observatory (CSO), allowed a conversion factor from
$\tau_{\rm 225~GHz}$ to $\tau_\lambda$ to be determined.  The
absolute flux scale was established from observations of Uranus,
Mars, Saturn, NGC2071, and CRL618 (Sandell 1994, 1998), and the
pointing and focus were checked regularly using calibrators near our
program sources.  SCUBA's 1.3~mm photometric pixel was also used to
make a strip scan perpendicular to the outflow for each object. The
uncertainty in the absolute flux calibration at 1.3~mm, 850, 750,
450, and 350~$\mu$m is 6, 8, 13, 20 and 30\% respectively.

SCUBA is located at a Nasmyth focus of the JCMT, without a beam
rotator, and images in equatorial coordinates have been produced
using a Gaussian interpolation function.  The arrays contained
several noisy or bad bolometers, which have been removed from the
data.  To avoid obtaining holes in the resulting images an
individual source was typically observed with between four and seven
different position offsets, and the data co-added in the final
stages of the data reduction after aligning the peaks in each
separate image.  Shifts of less than 2~arcsec were applied during
the alignment process, but this technique ensured that pointing
offsets between different integrations were minimized.  The typical
rms noise achieved at 850~$\mu$m was 20~mJy~beam$^{-1}$, and at
450~$\mu$m was 150~mJy~beam$^{-1}$.

The beam was measured at each wavelength by imaging Uranus (in 1997)
and Mars (in 1998), both of which had disk diameters of
approximately 4~arcsec at the time of observation.  The images of
Mars at 850, 750, 450 and 350~$\mu$m are shown in
Fig.~\ref{mars-fig}.  They illustrate that the beam is elongated in
the chop direction, even at 850~$\mu$m, although it is the
wavelength least affected by this problem.  They also show
significant power in the error beam, especially at the short
wavelengths. Because the images of the program sources were
combinations of several integrations, sometimes obtained on
different nights and therefore at different parallactic angles, only
the azimuthally-averaged radial intensity profile can be established
with any accuracy.  The beam at each wavelength has therefore been
fitted with the sum of two circular Gaussians
(Table~\ref{beam-fits}), more detailed modeling of the beam not
being justified by the data.  This produces excellent fits to the
beam profiles, and demonstrates that within a radius of 60 arcsec
80\% of the total power is in the central diffraction spike at
850~$\mu$m, reducing to 40\% at 350~$\mu$m.  The beam was not
measured at 1.3~mm, but a 22-arcsec Gaussian (FWHM) has been assumed
when comparing the 1.3~mm photometry with data at shorter wavelengths.

\section{Results}

\subsection{Images}
\label{results-images}

Fig.~\ref{scuba-images} displays the SCUBA images of the eight
protostars.  The greyscale is linear and spans the whole range of
emission, while the contours are logarithmic to emphasize the faint,
extended structure.  An arrow in the 850~$\mu$m image shows the
orientation of the blueshifted outflow for each source.  A glance at
the four protostars in Taurus shows immediately a trend between the
spatial distribution of the submillimeter dust emission and the
spectral classification of the objects: the Class~0 source, L1527,
comprises a central peak along with a considerable extended
envelope.  The Class~I protostars L1551~IRS5 and TMC1A have
comparable fractions of emission in an extended component compared
with the central peak, in spite of the fact that the luminosity of
these two sources differs by a factor of ten.  The emission from the
Class~II source HL~Tau is more compact than the other three sources,
although there remains some extended emission even around this object.

The images of the protostars in Perseus illustrate well the
different mode of star formation taking place in this cloud, which
contains many sources a factor of ten more luminous than those in
Taurus.  The youngest protostar of the sample is probably HH211-mm
(see Section~\ref{hh211} below), and like L1527, the dust emission
from this object is very extended with the envelope dominating over
any compact source.  The L1448 field contains not only the Class~0
protostar L1448-mm (also known as L1448C), but also several sources
associated with the bright IRAS source IRS3 to the north.  IRS3 is
now known to comprise a Class~0 binary system, L1448N, which appears
as an elongated source at position angle $\sim 45^\circ$ in
Fig.~\ref{scuba-images}, along with another Class~0 companion
20~arcsec away from the IRAS position, L1448NW (Terebey \& Padgett
1997; Barsony et al.\ 1998).  Approximately 30~arcsec to the
southeast of the peak submillimeter emission from NGC1333~IRAS2 a
second embedded object is detected at all wavelengths (see also
Sandell et al.\ 1994).  In order to distinguish the main bright peak
from this second source we designate them IRAS2:CR1 and IRAS2:CR2
respectively.  The ridge of dust emission from SVS13 was first
resolved into three protostars by Chini et al.\ (1997) at 1.3~mm and
referred to as MMS1, MMS2, and MMS3.  The same sources are now also
observed at submillimeter wavelengths in Fig.~\ref{scuba-images}. 
In the 850~$\mu$m image of SVS13 diamonds show the positions of the
water masers found by Haschick et al.\ (1980), and the crosses in
the SVS13 images at all wavelengths give the positions of the VLA
sources reported by Rodr\'\i guez et al.\ (1997).

\subsection{Continuum spectra}

Flux densities integrated within a radius of 45~arcsec for all the
program sources are listed in Table~\ref{fluxes}, and are plotted as
spectra along with measurements at other wavelengths available from
the literature in Fig.~\ref{spectra}.  It is from the spectra in
Fig.~\ref{spectra} that the luminosities listed in
Table~\ref{source-sample} have been derived.  They suggest that the
750~$\mu$m fluxes are consistently too low compared with the other
submillimeter measurements, by approximately 10--15\%.  While this
is within the 1$\sigma$ calibration uncertainties the fact that
every source is systematically low at 750~$\mu$m suggests some other
problem with the calibration which we have not been able to
identify, in spite of checking these data extremely carefully.
Alton, Davies, \& Bianchi (1999) also report a similar problem with
750~$\mu$m SCUBA data, although to a much greater extent than that
found here.

Table~\ref{fluxes} and Fig.~\ref{spectra} also give flux densities
for other sources detected in our SCUBA images.  Flux densities for
both L1448N and L1448NW are plotted separately, and the
NGC1333~IRAS2 region has been further separated into IRAS2:CR1 and
IRAS2:CR2.  Submillimeter dust emission from the SVS13 region
originates from the ridge containing MMS1, MMS2, and MMS3, and the
flux densities listed in Table~\ref{fluxes} are derived from CLEANed
images as described in Section~\ref{svs13}.  They are flux densities
per 11-arcsec beam, for comparison with the 1.3~mm measurements
given by Chini et al.\ (1997).

The extended emission associated with L1527 is not resolved into a
binary companion as reported by Fuller, Ladd, \& Hodapp (1996), but
instead is a continuous, low-level distribution tracing three arms
of a cross.  The emission from L1527 has therefore not been
separated into multiple components, and Table~\ref{fluxes} and
Fig.~\ref{spectra} give only integrated flux densities for this
source.  The HL~Tau field also covers the position of the T~Tauri
star XZ~Tau, 23~arcsec to the east.  The only wavelength for which
there is a statistically significant detection of submillimeter
continuum emission at the position of XZ~Tau is 850~$\mu$m, but
almost half of the 70~mJy~beam$^{-1}$ measured in this image is
caused by the error beam associated with HL~Tau, leaving a 3$\sigma$
upper limit for the 850~$\mu$m flux density of XZ~Tau of
48~mJy~beam$^{-1}$.  Assuming a spectrum $F_\nu \propto \nu^3$ for
any disk component associated with XZ~Tau, the 3$\sigma$ limits
obtained from the rms noise in each image (Table~\ref{fluxes})
correspond to $\sim 10$ to 20~mJy~beam$^{-1}$ at a wavelength of
1.3~mm, similar to the limit measured by Beckwith et al.\ (1990).

\section{Radial profile analysis}

\subsection{Description of models}

The major advantage of using dust emission for analyzing the
structure of protostellar envelopes derives from its simple
radiative transfer.  At submillimeter wavelengths the emission is
optically thin, and the dust opacity is a power-law function of
frequency, typically written as $\kappa_\nu = \kappa_0
(\nu/\nu_0)^\beta$, where $\beta$ lies in the range 0 to 2 depending
on the grain size, shape, and composition (e.g., Miyake \& Nakagawa
1993; Pollack et al.\ 1994; Agladze et al.\ 1994, 1996).  Although
$\beta$ is not well known, we will demonstrate that this is not a
significant problem for deriving density profiles from the dust
emission.  The absolute value of the dust opacity is also uncertain
to factors of 5 or so, but will primarily result in highly uncertain
masses derived from the dust emission, and does not influence the
derived density distributions.

The model assumes a spherically symmetric envelope, and the
intensity distribution of the dust emission in the SCUBA images is
used to derive the radial density profile in the envelope.  The
density profile is a power law, $\rho \propto r^{-p}$, the value of
$p$ derived using this method depending upon the assumed temperature
distribution.  For a power law temperature profile, $T \propto
r^{-q}$, this dependence of $p$ on the form of $T(r)$ is easily
demonstrated for the extreme limit of low optical depth, in the
Rayleigh-Jeans regime.  In this limit, such power-law profiles will
result in the intensity distribution of the dust emission, as a
function of angular distance on the sky, having the form
\begin{equation}
I_\nu \propto \theta^{1-(p+q)}.
\label{i-profile}
\end{equation}
Values of $p$ derived from $I_\nu(\theta)$ will therefore be
sensitive to the value of $q$ adopted.  The model uses the
analytical expression for the temperature distribution,
\begin{equation}
T(r) = 60 \left(\frac{r}{2 \times
10^{15}~{\rm m}}\right)^{-q} \left(\frac{L_{\rm bol}}{10^5~{\rm
L}_\odot}\right)^{q/2}~~~{\rm K},
\label{CBM98}
\end{equation}
where $q = 2/(4+\beta)$.  This arises from balancing the heating and
cooling of a dust grain at radius $r$ from a source of luminosity
$L_{\rm bol}$ (Chandler et al.\ 1998), and is a generalization of
the expressions derived by Scoville \& Kwan (1976), normalized to
match the more detailed radiative transfer calculations of Wolfire
\& Churchwell (1994).  Equation~\ref{CBM98} is assumed to hold until
the temperature falls to the temperature of the ambient cloud.  The
value of $\beta$ is not precisely known, but the 850 to 450~$\mu$m
spectral indices, $\alpha$ (where $F_\nu \propto \nu^\alpha$), lie
between 2.6 and 3.1 for all the sources.  Including a correction for
the shape of the Planck curve it is therefore likely that $\beta$ is
in the range 1 to 2, so temperature profiles which follow
Equation~\ref{CBM98} with $\beta = 1$ ($q = 0.4$) to a minimum
ambient temperature of $T_{\rm amb} = 10$~K, and one with $\beta =
2$ ($q = 0.33$) down to $T_{\rm amb} = 14$~K, are taken as two
extremes of possible temperature distributions.

In view of the uncertainty in the orientation on the sky of the
error beam of the telescope, particularly after several integrations
have been co-added, only azimuthally-averaged radial intensity
profiles are considered in deriving envelope density distributions
from the submillimeter dust emission.  The radial intensity profile
for each source is fitted by a combination of the emission from an
envelope, as described above, and a disk, which on the scale of
these measurements is a point source.  The envelope emission at
submillimeter wavelengths is assumed to be optically thin
throughout, and the best-fit models verify that this assumption is
valid.  The Rayleigh-Jeans approximation is {\it not} assumed to
hold, and corrections for the shape of the Planck curve are included
in the model.  The intensity profiles are fitted by varying the
index of the radial density power law, $p$, and the fraction of the
flux originating in the envelope, $f_{\rm env} = F_{\rm env} /
(F_{\rm env} + F_{\rm disk})$, where $F_{\rm env}$ is the envelope
flux and $F_{\rm disk}$ is the flux in the central point source
(disk). Clearly, the value of $F_{\rm env}$ (and therefore $f_{\rm
env}$) will depend upon the outer radius assumed for the envelope. 
Emission is present in the images even on the largest scales, with
no obvious outer radius for the envelopes evident in the data. Since
a chop throw of 120~arcsec is used, the envelopes have been modeled
with radial angular extents $\theta_{\rm o} = 60''$. However, the
effect of larger outer radii on the derived value of $p$ has also
been investigated by fitting the data with models having
$\theta_{\rm o} = 90''$ as well.  The inner radius is fixed at
100~AU, but the results are insensitive to its precise value since
these small radii contribute very little to the envelope flux at
these wavelengths.

The model intensity profile of the combined disk/envelope system is
computed for given values of $p$ and $f_{\rm env}$, it is convolved
with the model beam (Table~\ref{beam-fits}), and a DC offset with
the value of the convolved model at $\theta = 60''$ is subtracted to
simulate observation with the 120-arcsec chop throw.  Since
structure remains in the images at position angles perpendicular to
the chop direction, the same procedure of subtracting a DC offset
from the data was carried out before calculating the radial profiles
for comparison with the models.

\subsection{Modeling results}

Most of the sources exhibit relatively simple structure in their
dust emission, and a symmetrical envelope is not an unreasonable
model.  An obvious exception is the case of SVS13, where the narrow
ridge contains three separate protostars and a spherical envelope
model is clearly inappropriate.  No attempt has therefore been made
to model the emission from SVS13.  With one exception the images
used for the radial profile fitting are those displayed in
Fig.~\ref{scuba-images}.  For NGC1333~IRAS2:CR1, however, it was
clear that the nearby source IRAS2:CR2, and possibly the extension
to the northwest, might affect the profile analysis.  These two
regions of emission have therefore been selectively blanked before
forming the azimuthally-averaged profile.

Both the data and the model are averaged into 3-arcsec bins, and the
$\chi^2$ is calculated for a given value of $p$ and $f_{\rm env}$
using the statistical uncertainty at each data point.  The value of
$f_{\rm env}$ is not necessarily the same at each wavelength, and
depends on the relative values of the spectral index of the envelope
and disk components as a function of wavelength.  To avoid
introducing further free parameters into the modeling each source
and each wavelength have therefore been fitted separately.  A range
of $p$ from 0 to 5 is covered, and of $f_{\rm env}$ from 0 to 1.
These $\chi^2$ surfaces are then used to evaluate confidence limits
for $p$ and $f_{\rm env}$.  Fig.~\ref{profile-fits} illustrates the
results for individual sources by plotting the 850 and 450~$\mu$m
azimuthally-averaged radial profiles.  Data are shown along with the
model beam and the best-fit model profiles for the case where $\beta
= 1$ ($q = 0.4$), $T_{\rm amb} = 10$~K, $\theta_{\rm o} = 60''$. 
The contours enclose the 68.3, 95.4, and 99.7\% confidence limits in
the $\chi^2$-surfaces as a function of $p$ and $f_{\rm env}$.  In
general, the fitted value of $p$ is positively correlated with
$f_{\rm env}$.  Thus models with a steep density gradient in the
envelope, where most of the flux originates in the extended
component, may fit just as well as those where the value of $p$ is
lower, but where more flux comes from the disk. This result is
particularly clear for the Class~I source TMC1A, and the Class~II
object HL~Tau, both of which have very centrally-peaked profiles. In
these cases the model does not distinguish well between a large
value of $p$ and a high $f_{\rm env}$, and low $p$, low $f_{\rm env}$.

For each source the overall shape of the $\chi^2$ surface is similar
at all wavelengths, an encouraging sign that the model beams used
are reasonable.  The absolute value of the $\chi^2$ is probably not
very meaningful, since systematic uncertainties in the beam shape
have not been included in the calculation of the statistical
uncertainties.  Nevertheless, relative values are useful for
inter-comparison between sources.  The strongest contraints on $p$
and $f_{\rm env}$ are obtained from the 850~$\mu$m data, where the
signal-to-noise ratio is the highest.  Since the point spread
function at 850~$\mu$m is also less likely to be affected by
asymmetries introduced by the chop throw and by the error beam
(Fig.~\ref{mars-fig}), and the emission is the most optically thin
of all the wavelengths modeled, we conclude that the results
obtained from fitting the 850~$\mu$m profiles are the most
reliable.  The best 850~$\mu$m fits are found for HH211-mm, although
good fits are also obtained for TMC1A, L1527, and L1448-mm.  The
best fits for L1551~IRS5, HL~Tau, and NGC1333~IRAS2:CR1 are not
formally ``good'' fits based on the statistical uncertainties.

In order to examine possible evolutionary traits we would ideally
like to compare the fitted values of $p$ and $f_{\rm env}$ with a
continuously varying function of protostellar age, such as the
bolometric temperature, $T_{\rm bol}$, proposed by Myers \& Ladd
(1993).  Unfortunately no far-infrared measurements are available to
constrain the short-wavelength turnover for the spectrum of
HH211-mm, so its $T_{\rm bol}$ cannot be calculated.  Nevertheless,
the spectra in Fig.~\ref{spectra} have been used to calculate
$T_{\rm bol}$ for the other protostars, and assuming HH211-mm to be
the youngest, the sources may be ranked according to age within a
given spectral class.  Fig.~\ref{best-p-fenv} summarizes the results
of the profile fitting by plotting the best-fit values of $p$ and
$f_{\rm env}$, along with error bars corresponding to 95\%
confidence limits, as a function of spectral class within which our
sources are ranked as described above.  Only results for the
850~$\mu$m data and $\theta_{\rm o} = 60''$ are shown, but the
temperature profiles corresponding to both $\beta = 1$ and $\beta =
2$ are included.

For many of the sources the best-fit values of $p$ derived from the
850, 750, 450, and 350~$\mu$m images tend to decrease with
decreasing wavelength.  However, because the dynamic range in the
750, 450, and 350~$\mu$m images is lower than that at 850~$\mu$m,
the ranges of $p$ and $f_{\rm env}$ which give acceptable fits to
the data are larger.  The value assumed for $\theta_{\rm o}$ affects
both $f_{\rm env}$ and $p$. Higher values of $f_{\rm env}$ are
obtained for larger $\theta_{\rm o}$, for obvious reasons.  The
influence of $\theta_{\rm o}$ on $p$ is slightly more subtle, and
arises because of the subtraction of the 60-arcsec DC level.  This
truncation has the effect of steepening the output intensity profile
in the model for $\theta_{\rm o} = 90''$, and so values of $p$ lower
by approximately 0.1 are needed to fit the data compared with
$\theta_{\rm o} = 60''$.  From Equation~\ref{i-profile} it is clear
that the assumption of $\beta = 1$ compared with $\beta = 2$ will
give values of $p$ differing by at least 0.1.  In fact, the fits to
the measured intensity profiles show that $p$ derived assuming
$\beta = 1$ are approximately 0.2 lower than those for $\beta = 2$,
the extra 0.1 originating from the flattening of the temperature
profile at a higher ambient temperature for $\beta = 2$ in our model.

Fig.~\ref{best-p-fenv} shows that the 95\% confidence limits for $p$
encompass the values 1.5--2 expected for the collapse of
non-singular isothermal spheres, for most sources.  The large
uncertainty in $p$ for TMC1A and HL~Tau arises because of the low
$f_{\rm env}$ in these sources.  However, the decrease in $p$ from
values of around 2 in Class~0 objects to 1.5 in Class~I and II
sources predicted by the non-singular collapse solutions is not
observed.  Indeed, it is the protostars believed to be the youngest
in our sample which appear to have the most shallow density profiles
in their envelopes.  This result is made all the more extreme when
the values of $f_{\rm env}$ obtained from the model are compared
with those actually measured.  Some of the sources have had $F_{\rm
disk}$ determined directly through observations using the
interferometric link between the JCMT and CSO (Lay et al.\ 1994;
Brown et al.\ 1999).  Others have interferometry measurements of the
disk flux at millimeter wavelengths (Guilloteau et al.\ 1992;
Chandler, Carlstrom, \& Terebey 1994; Hogerheijde et al.\ 1997;
Gueth \& Guilloteau 1999), which can be extrapolated to 850~$\mu$m
assuming reasonable extremes of the spectral index of the disk
emission at millimeter/submillimeter wavelengths to be 2 and 3
(Beckwith \& Sargent 1991).  The range of ``measured'' values for
$f_{\rm env}$ calculated using these disk fluxes and the integrated
850~$\mu$m flux densities in Table~\ref{fluxes}, taking calibration
uncertainties into account, are plotted in Fig.~\ref{best-p-fenv}.
They agree with the 95\% confidence limits derived from the model
fitting, but are typically at the low end of the range.  The
correlation between $f_{\rm env}$ and $p$ (Fig.~\ref{profile-fits})
demonstrates that low $f_{\rm env}$ requires low $p$ to fit the
data, suggesting that the values of $p$ for HH211-mm and L1527 may
be even lower than 1.5.

The only models which predict values of $p$ less than 1.5 during
collapse are for cores which are initially highly centrally
condensed, and which evolve in a manner similar to the Shu solution
for the SIS\@.  In these models the shallow density profile occurs
immediately interior to the collapse expansion wave, and the size
scale on which the shallow density profile is observed should be
directly related to the age of the source via the sound speed, $a$. 
The ages of HH211-mm and L1527 are very uncertain, but estimates may
be obtained from the dynamical timescales of their outflows, which
are $\la 10^3$~yr and $\sim 10^4$~yr respectively (Gueth \&
Guilloteau 1999; Hogerheijde et al.\ 1998).  The age of L1527 has
also been estimated to be less than $10^4$~yr from modeling of its
SED in terms of the Shu picture by Kenyon, Calvet, \& Hartmann
(1993).  For sound speeds $a \sim 0.3$~km~s$^{-1}$ the radii of the
expansion waves for HH211-mm and L1527 are $\la 0.3''$ and $\sim
5''$, both of which lie within the central beam at 850~$\mu$m.  The
values of $p$ derived for these sources therefore apply to the
structure of their respective cloud cores {\it outside} the
expansion wave, in which case a steep initial density distribution
is strongly ruled out.  It seems that current models of cloud
collapse have problems reproducing our results.

Within the 95\% confidence limits our model does not show any
evolutionary trends in $f_{\rm env}$, but the best-fit envelope mass
needed to reproduce the observed 850~$\mu$m flux density, $M_{\rm
env}$, does decrease with spectral class.  Fig.~\ref{env-mass}
illustrates this for $\theta_{\rm o} = 60''$, where the upper value
of each range plotted corresponds to $\beta = 2$, and the lower
corresponds to $\beta = 1$, assuming $\kappa_\nu =
0.01(\nu/10^{12}~{\rm Hz})^\beta$ m$^2$~kg$^{-1}$ (Hildebrand
1983).  The {\it measured} $f_{\rm env}$, on the other hand, clearly
does decreases with spectral class, and is primarily caused by a
decrease in $F_{\rm env}$ rather than an increase in $F_{\rm
disk}$.  These results demonstrate that while the orientation of the
embedded protostar/disk/outflow system might have some effect on the
shape of the SED it does not {\it dominate} it, in agreement with
the recent work of Chen et al.\ (1995) and Ladd, Fuller, \& Deane
(1998).

\subsection{Influence of model assumptions on $p$}

The models with which our data are compared are spherically
symmetric, as is required of collapse from an initial state of
hydrostatic equilibrium in the absence of rotation or other forms of
support, such as a magnetic field.  While these are probably
reasonable as first approximations it is clearly essential that the
effect of deviations from the assumptions made in the modeling are
examined carefully, and the influence of outflows and other
asymmetries on the derived values of $p$ assessed.

\subsubsection{Variable $\beta$}

It is possible that the value of $\beta$ may vary across the
envelopes of the protostars, as has been observed for the dense
cores in NGC2024 and Orion A (Visser et al.\ 1998; Goldsmith,
Bergin, \& Lis 1997; Johnstone \& Bally 1999); such a variation
would affect the temperature distribution derived assuming
Equation~\ref{CBM98}.  In order to determine whether this may be a
problem for our analysis the model beams in Table~\ref{beam-fits}
have been deconvolved from the images presented in
Fig.~\ref{scuba-images}, which have then been smoothed to have the
resolution of a 22-arcsec Gaussian.  The spectral index $\alpha$ has
then been calculated across the source in the 1.3~mm scan direction,
perpendicular to the outflows.  Any systematic changes in $\alpha$
must then be due either to temperature (deviation from the
Rayleigh-Jeans regime), optical depth, or $\beta$.  Fig.~\ref{alpha}
shows the value of $\alpha$ derived from the 1.3~mm data and each
wavelength/source combination separately.  For NGC1333~IRAS2 we
found that we were chopping onto emission from SVS13 during the
1.3~mm observations, and so this source has not been included. 
SVS13 itself was not scanned at 1.3~mm because the source was too
complicated for the envelope analysis.  

Fig.~\ref{alpha} illustrates that for HL~Tau, L1527, L1448-mm, and
L1551~IRS5 the spectral index is quite flat as a function of distance. 
There is, however, some evidence for a decrease in $\alpha$ towards
the protostar for HH211-mm, where the trend is in a similar sense to
that found for NGC2024.  This decrease could be either due to high
optical depth in the envelope (unlikely, especially at 850~$\mu$m
and 1.3~mm), or due to a decrease in $\beta$; since the core is
known to contain an embedded protostar the decrease is not likely to
be caused by lower temperatures at the center.  However, the average
difference between the value of $\alpha$ away from the protostar
compared with the center of the cloud core is only $\sim 0.3$, so
even if the effect is caused entirely by a change in $\beta$ as a
function of radius, the two values of $\beta = 1$ and $\beta = 2$
assumed for our modeling should cover all possible extremes.  The
last source for which $\alpha$ appears to be a function of position
is TMC1A, where it tends to decrease away from the central source. 
This is consistent with lower temperatures at these radii because of
its low $L_{\rm bol}$, an effect which is included in our model.

\subsubsection{Outflow cavities and anisotropic density distributions}

For speed of computation our model is spherically symmetric, which
allows considerable areas of $p$--$f_{\rm env}$ space to be explored
when fitting the data, but does not easily enable the inclusion of
the effects of an outflow cavity or other anisotropy in the density
distribution.  Several models have therefore been constructed
including conical outflow cavities, which can also be regarded as
extreme cases of flattened, anisotropic, density profiles in the
envelope.  The four distinct combinations of outflow opening angle
and inclination identified by Cabrit \& Bertout (1986) have been
examined, and are illustrated schematically in Fig.~\ref{cavity}.
The resulting normalized radial intensity profiles for the envelopes
are also plotted.  No central point source has been included, which
would only serve to decrease the difference between the various
cases as $f_{\rm env}$ decreases.  Cases 1, 2, and 3, have outflow
semi-opening angles of $30^\circ$, and case 4 has a semi-opening
angle of $60^\circ$.  Fig.~\ref{cavity} shows that for those cases
which have some fraction of the outflow cavity along the line of
sight to the central protostar, cases 1 and 4, the normalized radial
intensity profile is more shallow than an envelope with no cavity. 
Outflow cavities which lie away from the line of sight have profiles
slightly steeper, although they are extremely close to the profile
obtained for an envelope with no cavity.  Since none of the
protostars in our sample has case 1 or 4 outflows, the effect of not
including an outflow cavity in our model will result in fitted
density profiles which, if anything, may be slightly too steep.

\subsubsection{Shock heating of dust by the outflow}
\label{shock-heating}

The one effect which might contribute to the low $p$ derived from
our models for L1527 and HH211-mm is if the radial temperature
distribution, averaged over all solid angles, does not follow
Equation~\ref{CBM98} because of sources of heating in the envelope
other than radiation from the central protostar.  Shock heating at
the interface between the cloud and the outflow is demonstrated
below to be a possible explanation for the low $p$ derived for L1527
(Section~\ref{l1527}), but for HH211-mm the evidence for this is in our
submillimeter images is less convincing (partly due to its larger
distance).  There is, however, considerable shocked molecular
hydrogen emission associated with the HH211 outflow (McCaughrean,
Rayner, \& Zinnecker 1994), so shock heating of the dust cannot be
altogether discounted.  Nevertheless, the solid angle subtended by
the HH211 outflow, $\sim 2$\% of the whole envelope, implies that
the mass-weighted temperature of the shock-heated dust must be
signficantly greater than 50~K for it to increase the {\it mean}
temperature by more than 1~K.

Quantification of the influence of shocks on the mean dust
temperature profile of HH211-mm will require higher spatial resolution
measurements than those presented here, but the magnitude of the
effect can be estimated through the following argument.  The time
for a dust grain to cool from $\ga 1000$~K to a few tens of K
through radiation is very short, on the order of a day.  The average
dust temperature in the outflow will therefore be determined by
other processes, and provided the post-shock density is high enough
to couple the dust and gas via collisions, the cooling may actually
be dominated by molecular line emission.  The excitation
temperatures measured in high-velocity molecular gas may therefore
be good indicators of the dust temperature in the outflow, and in
some sources, the bulk of the high-velocity gas exhibits excitation
temperatures have been observed to be 10--20~K higher than the
surrounding medium (e.g., Umemoto et al.\ 1992; Chandler et al.\
1996).  This enhancement is also comparable to the color temperature
of 25~K derived from 60 and 100~$\mu$m dust emission for the L1551~IRS5
flow (Clark \& Laureijs 1986; Edwards et al.\ 1986).  However, it is
insufficient to affect significantly the mean temperature profile in
the envelope of HH211-mm.

\subsubsection{Uncertainties in the distance to Perseus}

The effect of an error in the assumed distance to the Perseus
sources has also been investigated, as a fairly wide range of values
is reported in the literature.  We find that the influence of $D$ on
the fitted profile is negligible, which can be seen from the
following reasoning.  From Equation~\ref{CBM98} the physical radius
at which the dust temperature has some fiducial value $T_0$ is
proportional to $L_{\rm bol}^{1/2}$, so the corresponding angular
radius is proportional to $L_{\rm bol}^{1/2}/D$.  Since the assumed
source luminosity is proportional to $D^2$, the angular radius
corresponding to $T_0$ is independent of $D$.  The fitted values of
$p$ and $f_{\rm env}$ are thus unaffected by $D$ because the outer
radius of the envelope is defined in terms of angular distance,
there being no obvious edge to the emission in the images.  The main
difference between models which assume a different value of $D$ is
at small radii, where the inner radius of the envelope is fixed at
100~AU rather than at some value of $\theta$.  However, the flux
from these inner regions originates typically from less than
1~arcsec in angular extent for values of $D$ between 200 and 350~pc,
and so contributes little to the central 10--15~arcsec beam.

\section{Structure of individual sources}

Extended structure related to the local environment of each source,
and possibly the interaction of the source with the surrounding
cloud, clearly influences the observed submillimeter emission in
Fig.~\ref{scuba-images} at some level.  Features particular to each
protostar are described in more detail below, and we discuss how
they might affect the derived density profiles.

\paragraph{L1527}
\label{l1527}

The extended dust emission from L1527 delineates three arms of a
cross.  This structure must be due to a combination of increased
column density and/or temperature, and coincides with edges of the
outflow cavity defined by the CO(3--2) emission (MacLeod et al.\
1994; Hogerheijde et al.\ 1998).  To investigate the relative
contributions of density and temperature to the observed
submillimeter emission our dust maps have been compared with the
high-resolution HCO$^+$(1--0) interferometric image of Hogerheijde
et al.\ (1998).  We find that the arms of the dust emission coincide
very closely with cross features in the HCO$^+$.  This indicates
that at least some of the dust features are caused by a column
density enhancement along the edges of the outflow, and perhaps
trace a shell of gas and dust swept-up by the driving wind.

The 450~$\mu$m image (Fig.~\ref{scuba-images}) has also been
smoothed to match the resolution at 850~$\mu$m using the model beams
in Table~\ref{beam-fits}, and used to produce a map of the dust
color temperature, $T_{450/850}$, in regions which have a
sufficiently high signal-to-noise ratio.  Fig.~\ref{l1527-fig}
illustrates the resulting map as a greyscale, overlaid by the
interferometric HCO$^+$ contours.  Assuming a value for $\beta$ of
1.5 and optically-thin emission, the color temperature is between 10
and 40~K throughout the envelope.  However, it clearly is not
decreasing monotically with distance away from the central
protostar.  A peak in the local dust temperature is located along
the northeast arm of the HCO$^+$(1--0) cross, at the point where
there is a corresponding decrease in HCO$^+$ emission.  The opposite
effect can be seen at the end of the southeastern arm, where a small
HCO$^+$ peak coincides with a local minimum in $T_{450/850}$.  One
possible explanation for this anti-correlation between the
HCO$^+$(1--0) and $T_{450/850}$ is that the partition function of
HCO$^+$ increases with increasing temperature, suppressing the
emission in the $J$=1--0 transition relative to surrounding regions
of cooler gas. The coincidence of local temperature peaks with the
edges of the outflow cavity lead us to propose that they are caused
by shocks, a hypothesis which could be tested through observations
of infrared H$_2$ emission.

The excitation temperature derived for the high-velocity CO emission
from the L1527 outflow is 45~K, a significant enhancement over the
temperature of the bulk of the molecular gas, 25~K (Hogerheijde et
al.\ 1998).  By an argument similar to that given for HH211-mm in
Section~\ref{shock-heating}, the large opening angle of the L1527
flow means that shocks might be an important source of heating away
from the protostar, flattening the radial temperature profile in the
envelope relative to that possible from radiative heating alone. 
This might, to some extent, explain our low value of $p$.

\paragraph{L1551~IRS5}

The distinct cross-shaped structure in the submillimeter dust
emission from L1551~IRS5 first reported by Ladd et al.\ (1995) is
confirmed by our SCUBA images.  These authors also suggested that
the origin of the cross feature lies in the interaction of the
outflow with the surrounding cloud core, and comparison with the
morphology of 2~$\mu$m reflection nebulosity (e.g., Hodapp et al.\
1988; Lucas \& Roche 1996) shows that the orientation of structure
in the outflow cavity extends to arcsecond scales.  Any contribution
to the heating of the envelope around IRS5 by its outflow will tend
to result in an underestimate of $p$.  To the northeast of IRS5 the
dust emission follows a ridge of molecular gas extending towards
L1551NE.  

\paragraph{TMC1A}

Structure in the dust envelope surrounding TMC1A corresponds well to
the molecular core in interferometer maps of $^{13}$CO(1--0)
emission (e.g., Brown \& Chandler 1999).  While a large range of $p$
fits the radial profiles reasonably well, this source is one for
which the shape of the dust SED is best explained by $p \sim 1$
(Chandler et al.\ 1998).  The SED modeling was carried out using
flux densities integrated over the relatively small maps which
single-element detector systems were able to produce, and so apply
to the envelope close to the source.  The radial profile modeling
is, on the other hand, more dominated by the points at larger radii,
which have lower statistical uncertainties compared with those at
the central pixel.  It is notable that the observed profile is more
shallow than the best-fit 850~$\mu$m profile shown in
Fig.~\ref{profile-fits} at small radii.  Furthermore, the measured
$f_{\rm env}$ for TMC1A places the most probable value of $p$
towards the lower end of the distribution plotted in
Fig.~\ref{best-p-fenv}.  Thus while a precise estimate of $p$ for
the envelope around TMC1A cannot be obtained from the radial profile
fitting because of the relatively low fraction of the flux
originating in the envelope, a combination of measurements now
suggests that its radial density profile is quite flat close to the
protostar.  This is consistent with the SIS collapse models for a
protostar $\sim {\rm few} \times 10^4$~yr old, but also the
kinematics of the $^{13}$CO(1--0) emission show that the envelope is
in approximate Keplerian rotation (Ohashi et al.\ 1997; Brown \&
Chandler 1999), which might provide the support necessary to explain
a value of $p < 1.5$ for non-singular initial conditions.

\paragraph{HL~Tau}

A faint envelope is detected even around the Class~II protostar
HL~Tau.  Since the model is not a particularly good description of
the source we have investigated whether outer radii of $\theta_{\rm
o} = 30''$ fit the radial profiles better than the 60$''$ and 90$''$
used for the younger sources, but find it does not.  The low level
structure is consistent with that expected for a remnant
circumstellar envelope most of which has already been dispersed by
the outflow or accreted onto the central protostar.  The north-south
extension evident in Fig.~\ref{scuba-images} can be traced through
molecular line emission to within a few arcsec of the source (Cabrit
et al.\ 1996).

\paragraph{HH211-mm}
\label{hh211}

Comparison of the submillimeter continuum emission from HH211-mm with
the H$^{13}$CO$^+$(1--0) image of Gueth \& Guilloteau (1999) shows
good agreement between the dust distribution and the structure of
the dense gas.  At the high resolution of 350 and 450~$\mu$m
(Fig.~\ref{scuba-images}) an extension along the direction of the
redshifted outflow lobe is also observed.  The flow in this source
is particularly highly collimated and jet-like, but its larger
distance and the resulting low linear resolution of the 850~$\mu$m
data preclude a color temperature analysis similar to that carried
out for L1527.  The dynamical timescale for the molecular jet is
less than $10^3$~yr (Gueth \& Guilloteau 1999), suggesting the
protostar is extremely young.

The semi-opening angle of the outflow cavity at the base of the jet,
projected onto the sky, is approximately 15$^\circ$.  This provides
an upper limit to the true opening angle, and as Fig.~\ref{cavity}
and the discussion in Section~\ref{shock-heating} illustrate,
suggests that the influence of the outflow on the derived value of
$p$ is minimal for this source.

\paragraph{L1448-mm}

The envelope around L1448-mm is the most circularly symmetric of our
Perseus objects, so radiative heating by the protostar is expected
to dominate the temperature distribution, with shocks having a
relatively small effect on the value of $p$ derived for this
source.  However, some faint, extended structure at 850~$\mu$m is
observed towards the north, in the direction of the blueshifted
outflow lobe.  This feature is also detected in the HIRES-processed
IRAS data by Barsony et al.\ (1998), who suggest that the extended
far-infrared emission is associated with the collision of the
L1448-mm outflow with an outflow from one of the binary companions
comprising L1448N\@.  The reader is directed to the paper by Barsony et
al.\ (1998) for a recent, more detailed description of star
formation in the L1448 cloud.

\paragraph{NGC1333~IRAS2}

The 1.25~mm dust emission from NGC1333~IRAS2 was found to be
elongated towards the southeast in the IRAM 30-m images presented by
Lefloch et al.\ (1998), but was not resolved into the individual
sources IRAS2:CR1 and IRAS2:CR2.  IRAS2:CR2 was, however, detected
as a separate peak at 800~$\mu$m by Sandell et al.\ (1994).  The
presence of two distinct outflows originating close to IRAS2 led
Sandell et al.\ to propose that the cloud core harbours a binary
system, although molecular line data with sufficiently high
resolution are not available for us to determine the relationship
between IRAS2:CR2 and these outflows.  The complicated structure of
the cloud, combined with multiple flows, make the interpretation of
the extended emission from NGC1333~IRAS2 in terms of cloud/wind
interactions too difficult, and must await more detailed observations.

\paragraph{SVS13}
\label{svs13}

The elongated ridge associated with the infrared source SVS13 was
mapped with high resolution in 1.1~mm and 800~$\mu$m continuum
emission by Sandell et al.\ (1990), and has subsequently been
resolved into three distinct millimeter sources by Chini et al.\
(1997).  These objects, designated MMS1 (SVS13), MMS2 (also known as
SVS13B), and MMS3 by Chini et al., have also been identified
interferometrically (Grossman et al.\ 1987; Bachiller et al.\
1999).  SVS13B drives an energetic molecular jet (Bachiller et al.\
1999), and MMS3 is coincident with radio continuum sources (Rodr\'\i
guez et al.\ 1997) and a water maser (Haschick et al.\ 1980).  The
resolution of our 350 and 450~$\mu$m SCUBA images is sufficient to
distinguish MMS1, MMS2, and MMS3, along with extended emission
outlining the outflow cavity towards the southeast (see also Lefloch
et al.\ 1998).  Our 750 and 850~$\mu$m images do not, however,
resolve these sources, and for a more detailed comparison of their
spectral energy distributions and determination of their
evolutionary states, higher resolution is needed.  We have therefore
attempted to deconvolve all our SVS13 images using a CLEAN algorithm
and our model beams, and have produced maps with the resolution of
an 11-arcsec (FWHM) Gaussian to match the 1.3~mm measurements of
Chini et al.  This process was very successful at 750 and
850~$\mu$m, for which the telescope beam of the JCMT is most
symmetric, and the result for 850~$\mu$m is illustrated in
Fig.~\ref{svs13-clean}.  

Flux densities per 11-arcsec beam are plotted for MMS1, MMS2, and
MMS3, in Fig.~\ref{spectra}.  The absolute errors are dominated by
the calibration uncertainties, but comparison between sources can be
made with much higher accuracy than implied by the error bars.  It
can then be seen that there is a trend for the submillimeter
spectral index to decrease from MMS1 to MMS2 to MMS3, consistent
with these latter sources having lower luminosities and/or being
more deeply embedded.  High resolution far-infrared measurements
will be needed to constrain the SEDs of these objects further, but
it seems likely that MMS2 and MMS3 are Class~0 protostars.

\section{Discussion and conclusions}
\label{discussion}

We have obtained the best constraints yet on the density profile in
protostellar envelopes on scales of a few 1000 AU, and find that a
single radial power law for the density is a good fit to the
intensity profile of the dust emission.  For most sources the
observations are consistent with $1.5 \la p \la 2$, as expected for
the collapse of an isothermal sphere in an initial (non-singular)
equilibrium configuration.  There are, however, some notable
exceptions.  In particular, the value of $p$ derived for the
youngest sources in our sample, the Class~0 protostars HH211-mm and
L1527, is considerably lower than would be expected from the models
described in Section~\ref{introduction}, which predict $p \sim 2$ at
early times.  The density profile may also be relatively shallow for
the Class~I source TMC1A, but this may be explained either by the
shallow profile expected inside the expansion wave for the collapse
of highly centrally condensed cloud cores, or by the fact that the
protostar has already accreted most of the low-angular momentum
material from its envelope, leaving behind material which is
centrifugally supported.  For the Class~0 protostars, however, other
explanations are needed for low $p$.  While there are obvious
problems with our radial profile analysis, such as the fact that we
do not include an outflow cavity, and that the outer radius of the
envelope may be larger than 60~arcsec, most of these simplifications
result in $p$ being overestimated, thus accentuating the shallow
density distributions.

In the case of L1527 heating due to the interaction of a wind from
the protostar with the surrounding envelope may enhance the
temperature over that expected from radiative heating alone, and
might account for the exceptionally low value of $p$ derived from
our modeling.  Indeed, the ubiquity of extended dust emission
associated with the flows from the protostars in our current sample
indicates that submillimeter dust emission may prove to be extremely
useful for investigating jet/cloud interactions in other very
deeply-embedded protostellar systems.  For HH211-mm, however, the
solid angle subtended by the outflow is very small, and not even
outflow heating seems to be able to explain the intensity profile of
the submillimeter dust emission.  We therefore believe the shallow
density distribution obtained for HH211-mm to be a robust result.
Because HH211-mm is the youngest protostar in our sample, this has
important implications for models of cloud collapse.

At a distance of 350~pc, the envelope of HH211-mm is well fitted at
all radii by a single power law, $\rho \propto r^{-1.5}$, out to
0.1~pc.  Values of $p$ as high as 2 are strongly excluded.  The
outer parts of the dense cloud core may as yet retain the
distribution it had prior to collapse, and distributions flatter
than $\rho \propto r^{-2}$ then imply that the core is supported at
some level, perhaps by magnetic fields or turbulence.  In a picture
of star formation where collapse begins from an unstable equilibrium
configuration when all such forms of support are lost, the inner
parts should, however, have attained a distribution $\rho \propto
r^{-2}$.  For example, if the final mass of the star forming at the
centre is $\sim 1$~M$_\odot$, all the material within a radius of
$\sim 20$~arcsec will, at some stage, have to end up in the star,
and the density distribution within this radius would be expected to
reflect the $r^{-2}$ predicted by collapse theory.  Furthermore, if
the dynamical timescale of the outflow from HH211-mm is a rough
indicator of the age of this source, the radius of the infall region
in the SIS collapse picture is much less than 1~arcsec, ruling out
the possibility that the $r^{-1.5}$ density profile represents an
extended region of free fall unless the collapse did {\it not}
originate from an equilibrium configuration.  Another possibility is
that most of the cloud core remains supported and that the support
mechanism is lost from the inside out.  In this case the accretion
rate must be determined by the rate at which support is lost rather
than hydrodynamical properties of the envelope, such as the local
sound speed.  

Our SCUBA measurements demonstrate that the accuracy with which
density profiles can be established is no longer limited by the
dynamic range available in the single-dish images, but instead is
determined by other more systematic problems, such as uncertainties
in the beam shape.  Our use of azimuthally-averaged intensity
profiles was necessitated by the asymmetries in the JCMT beam, and
while more detailed, three-dimensional, radiative transfer modeling
might be desirable in the future, it is not yet justified by the
data.  Even within the current analysis the accuracy of measured
values of $f_{\rm env}$, which are needed to provide the strongest
constraints on $p$, is determined by the absolute calibration
uncertainties of data obtained at different telescopes through the
highly variable terrestrial atmosphere.  For Class~I/II sources,
where the contribution from a disk may be significant, the main
problem is therefore obtaining reliable relative calibrations for
single-dish and interferometer measurements.  Future instruments
such as the Smithsonian Submillimeter Array will be very important
in this respect, since its coverage of a wide range of spatial
scales will enable $f_{\rm env}$ to be determined directly.  From
the more general perspective of establishing more detailed
evolutionary trends in envelope structure the present work is
clearly limited by its small sample, and illustrates the need for
much larger surveys of entire star forming regions.

\acknowledgments

The authors are grateful to Ant Whitworth for useful discussions
about self-similar collapse solutions, and the anonymous referee for
comments which have clarified and improved the paper.  The JCMT is
operated by the Joint Astronomy Centre on behalf of the Particle
Physics and Astronomy Research Council of the United Kingdom
(PPARC), the Netherlands Organisation for Scientific Research, and
the National Research Council of Canada.  CJC and JSR acknowledge
the support of a PPARC Advanced Fellowship and a Royal Society
Fellowship respectively.

\newpage

\figcaption{\label{mars-fig}Linear greyscale images of Mars obtained
in 1998 January, at wavelengths of 850, 750, 450, and 350~$\mu$m.
Contours are logarithmic, and are drawn at 2, 4, 8, 16, 32, and 64\%
of the peak in each case.  The direction of the 120-arcsec chop
throw is marked in the 850~$\mu$m image.  The beam is clearly
extended in the chop direction even at 850~$\mu$m, and at 450 and
350~$\mu$m the beam comprises a central diffraction spike, and an
approximately triangular error pattern which becomes more symmetric
on large scales.}

\figcaption{\label{scuba-images}Submillimeter continuum images of
the eight protostars observed with SCUBA, at 850, 750, 450, and
350~$\mu$m.  The greyscale is linear and spans the full range of
emission.  The contours are logarithmic starting at a level of
2$\sigma$ and spaced by intervals of $\times 1.73$.  The rms noise
for all the sources at all wavelengths is listed in
Table~\ref{fluxes}.  An arrow in the 850~$\mu$m image, of length
60~arcsec for reference, shows the direction of the blueshifted
outflow for each source.  All images cover $220 \times
220$~arcsec$^2$.  The positions of the H$_2$O masers reported by
Haschick et al.\ (1980) are plotted as diamonds in the 850~$\mu$m
image of SVS13, and the VLA sources from Rodr\'\i guez et al.\
(1997) are shown as crosses. The dotted contours outline the edges
of the areas covered by SCUBA.}

\figcaption{\label{spectra}Continuum spectra for the eight
protostars in our sample.  For panels containing spectra for just
one source, the SCUBA flux densities, integrated within a radius of
45~arcsec, are shown as filled triangles.  Other measurements
obtained in large beams are open squares.  For the L1448 field flux
densities for the protostellar binary L1448N, and the Class~0
protostar L1448NW, are also plotted, derived as described in the
notes to Table~\ref{fluxes}.  In this case, and in other panels
containing spectra for multiple sources, separate symbols have not
been used for the SCUBA points, to avoid the plots becoming too
confused.  The integrated flux density from NGC1333~IRAS2 within a
45-arcsec radius is given along with the separate measurements for
the central protostar IRAS2:CR1, and the southeastern source
IRAS2:CR1 (Table~\ref{fluxes}).  Emission from the three protostars
in the SVS13 ridge has been separated by deconvolving the images as
described in the text, and are in units of Jy per 11-arcsec Gaussian
beam.  References for non-SCUBA flux densities are as follows.
L1527: far-infrared measurements have been obtained from
HIRES-processed IRAS images (unpublished).  L1551~IRS5:
near/mid-infrared data are from Cohen \& Schwarz (1983) and Beichman
\& Harris (1981); far-infrared data are from Davidson \& Jaffe
(1984) and Cohen et al.\ (1984); millimeter/submillimeter data are
from Ladd et al.\ (1995) and Walker, Adams, \& Lada (1990); the
3.4~mm measurement comes from Keene \& Masson (1990).  TMC1A:
references are given by Chandler et al.\ (1998).  HL Tau: infrared
measurements are from Strom et al.\ (1989); millimeter/submillimeter
data come from Weintraub, Sandell, \& Duncan (1989), Adams, Emerson,
\& Fuller (1990), Beckwith \& Sargent (1991), Beckwith et al.\
(1990), Sargent \& Beckwith (1991), Mundy et al.\ (1996), and Blake,
van Dishoeck, \& Sargent (1992).  L1448: all non-SCUBA measurements
come from Barsony et al.\ (1998).  NGC1333~IRAS2: far-infrared data
are from Jennings et al.\ (1987); millimeter measurements come from
Sandell et al.\ (1994) and Lefloch et al.\ (1998).  SVS13:
far-infrared data are from Jennings et al.\ (1987); 1.3~mm flux
densities per 11-arcsec beam for MMS1, MMS2, and MMS3 are from Chini
et al.\ (1997).  Where multiple measurements are reported for a
particular wavelength a weighted mean has been plotted.}

\figcaption{\label{profile-fits}{\it Left}: Normalized,
azimuthally-averaged, radial profiles for our protostars at 850 and
450~$\mu$m (points).  The best-fit model for $\theta_{\rm o} = 60''$
and $\beta = 1$ ($q = 0.4$) is also plotted as a solid line, and the
model beam profile from Table~\ref{beam-fits} is given as a dot-dash
line.  For NGC1333~IRAS2:CR1 the radial profile is derived from an
image blanked as described in the text.  {\it Right}: contours of
the 68.3, 95.4, and 99.7\% confidence limits as a function of $p$
and $f_{\rm env}$, with the best-fit values marked by a filled
triangle.}

\figcaption{\label{best-p-fenv}Best-fit values for $p$ and $f_{\rm
env}$ plotted along with 95\% confidence limits, as a function of
spectral class.  The results are derived from the 850~$\mu$m
profiles and are shown for models with $\theta_{\rm o} = 60''$. 
Temperature profiles which assume $\beta = 1$ ($q = 0.4$) are open
circles, those with $\beta = 2$ ($q = 0.33$) are filled circles. 
The range of values for $f_{\rm env}$ obtained directly from
measurements are indicated by grey bars.  The values of $p$
corresponding to the $r^{-2}$ of a singular isothermal sphere and
the $r^{-3/2}$ for free-fall collapse are also marked.}

\figcaption{\label{env-mass}Best-fit envelope masses, $M_{\rm env}$,
needed to account for the observed 850~$\mu$m flux density, as a
function of spectral class.  The lower value corresponds to $\beta =
1$, the upper to $\beta = 2$.}

\figcaption{\label{alpha}Spectral indices, $\alpha$, defined by
$F_\nu \propto \nu^\alpha$, between $\lambda = 1.3$~mm and the
wavelengths of the other SCUBA filters.  The values of $\alpha$ are
plotted as a function of position away from the central protostar,
along the 1.3~mm scan direction, perpendicular to the outflow from
each source.}

\figcaption{\label{cavity}Normalized radial profiles for envelopes
having conical outflow cavities, with opening angle and inclination
combinations as defined by Cabrit \& Bertout (1986).  The envelope
is the hatched area in all cases.  These examples are for
$\theta_{\rm o} = 60''$, and $f_{\rm env} = 1$.  The semi-opening
angle of the cavity is $30^\circ$ for cases 1, 2, and 3, and for
case 4 it is $60^\circ$.  The inclination angles of the outflow
symmetry axis to the line of sight are $0^\circ$, $45^\circ$,
$90^\circ$, and $45^\circ$ for cases 1, 2, 3, and 4 respectively, as
shown.}

\figcaption{\label{l1527-fig}Greyscale of the 450 to 850~$\mu$m dust
color temperature for L1527, assuming optically-thin emission and
$\beta = 1.5$.  The position of the protostar is indicated by a
star.  Contours of the HCO$^+$(1--0) emission from Hogerheijde et
al.\ (1998) are overlaid, illustrating the anti-correlation between
local enhancements in the dust temperature and molecular emission,
particularly along the eastern arms of the cross.}

\figcaption{\label{svs13-clean}Image of the 850~$\mu$m emission from
SVS13, after CLEANing with the model beam and reconstruction with an
11-arcsec Gaussian.  The positions of the millimeter continuum
sources measured using the Plateau de Bure interferometer by
Bachiller et al.\ (1999) are plotted as asterisks, and are labeled
with the names given to these objects by Chini et al.\ (1997).}

\begin{deluxetable}{lcccccc}
\small
\tablecaption{\label{source-sample}Source sample.}
\tablehead{
\colhead{Source} & \colhead{Cloud} & \colhead{R.A. (2000)} &
\colhead{Dec.\ (2000)} & \colhead{Ref.\tablenotemark{a}} &
\colhead{Class} & \colhead{$L$ (L$_\odot$)} \\
\colhead{} & \colhead{} & \colhead{(h~~~~m~~~~s)} &
\colhead{($^\circ~~~~'~~~~''$)} & \colhead{} & \colhead{}
}
\startdata
L1448-mm   & Perseus & 03 25 38.84 & 30 44 05.4 & 1 & 0 & 14 \nl
NGC1333~IRAS2 & Perseus & 03 28 55.59 & 31 14 37.5 & 2 & 0 & 35 \nl
SVS13      & Perseus & 03 29 03.75 & 31 16 03.7 & 3 & II & 50 \nl
HH211-mm & Perseus & 03 43 56.73 & 32 00 51.9 & 2 & 0 &
\hspace{-4mm}$\sim 15$\tablenotemark{b} \nl
L1551~IRS5 & Taurus & 04 31 34.15 & 18 08 05.2 & 4 & I & 21 \nl
HL~Tau     & Taurus & 04 31 38.41 & 18 13 57.8 & 5 & II & \phn7 \nl
TMC1A      & Taurus & 04 39 35.16 & 25 41 45.1 & 6 & I & \phn2 \nl
L1527      & Taurus & 04 39 53.88 & 26 03 10.2 & 7 & 0 & \phn2 \nl
\enddata
\tablenotetext{a}{References for the positions are as follows: (1)
Guilloteau et al.\ (1992); (2) this paper (for NGC1333~IRAS2 the
position is that of the submillimeter continuum peak, designated
IRAS2:CR1 in Section~\ref{results-images}); (3) Bachiller et al.\
(1999); (4) Keene \& Masson (1990); (5) Mundy et al.\ (1996); (6)
Brown \& Chandler (1999); (7) Chandler et al.\ (1994).  Positions
measured from the SCUBA images are accurate to approximately
3~arcsec.}
\tablenotetext{b}{Since only millimeter/submillimeter fluxes are
available for HH211-mm, its luminosity has been estimated by
comparing its spectrum with those of L1448-mm and NGC1333~IRAS2.}
\end{deluxetable}

\begin{deluxetable}{ccccc}
\small
\tablecaption{\label{beam-fits}Two-component Gaussian fits to the
JCMT beam.}
\tablehead{
\colhead{$\lambda$} & \colhead{$\theta_1$\tablenotemark{a}} &
\colhead{$P_1$} & \colhead{$\theta_2$} & \colhead{$P_2$} \\
\colhead{($\mu$m)} & \colhead{(arcsec)} & \colhead{} &
\colhead{(arcsec)} & \colhead{}
}
\startdata
850 & 13.8 & 0.78 & 47 & 0.22 \nl
750 & 12.4 & 0.72 & 43 & 0.28 \nl
450 & \phn8.3 & 0.53 & 37 & 0.47 \nl
350 & \phn7.9 & 0.41 & 31 & 0.59 \nl
\enddata
\tablenotetext{a}{FWHM ($\theta$) and fractional power ($P$) for
Gaussian components 1 and 2, given the Gaussian interpolation from
Nasmyth to equatorial coordinates.}
\end{deluxetable}

\begin{deluxetable}{lr@{$\pm$}lcr@{$\pm$}lcr@{$\pm$}lcr@{$\pm$}lc}
\small
\tablecaption{\label{fluxes}Integrated submillimeter flux densities
for the protostars listed in Table~\ref{source-sample}.}
\tablehead{
\colhead{Source} & \multicolumn{3}{c}{$\lambda$ = 850 $\mu$m} &
\multicolumn{3}{c}{750 $\mu$m} & \multicolumn{3}{c}{450 $\mu$m} &
\multicolumn{3}{c}{350 $\mu$m} \\
\colhead{} & \multicolumn{2}{c}{$F_{\rm int}$\tablenotemark{a}~~(Jy)}
& \colhead{Rms\tablenotemark{b}} &
\multicolumn{2}{c}{$F_{\rm int}$} & \colhead{Rms} &
\multicolumn{2}{c}{$F_{\rm int}$} & \colhead{Rms} &
\multicolumn{2}{c}{$F_{\rm int}$} & \colhead{Rms}
}
\startdata
L1448-mm & 5.34 & 0.43 & 19 & 6.91 & 0.91 & 30 & 34.2 & 6.9 & \phn98
& 58 & 18 & 490 \nl
L1448N & 6.53 & 0.52 & \nodata & 8.3 & 1.1 & \nodata & 43
& 12 & \nodata & 72 & 22 & \nodata \nl
L1448NW & 1.84 & 0.15 & \nodata & 2.60 & 0.34 & \nodata &
13.1 & 2.6 & \nodata & 19.5 & 5.9 & \nodata \nl
NGC1333~IRAS2 & 8.97 & 0.73 & 24 & 11.9 & 1.6 & 37 & 68 & 14 & 140 &
119 & 36 & 530 \nl
~~~~~~IRAS2:CR1 & 4.79 & 0.39 & \nodata & 6.61 & 0.86 & \nodata & 43 &
11 & \nodata & 74 & 22 & \nodata \nl
~~~~~~IRAS2:CR2 & 1.19 & 0.10 & \nodata & 1.53 & 0.21 & \nodata & 7.7
& 1.6 & \nodata & 13.4 & 4.3 & \nodata \nl
SVS13 & 14.9 & 1.2 & 36 & 21.5 & 2.8 & 61 & 119 & 24 & 200 & 203 &
61 & 850 \\
~~~~~~MMS1 & 2.68 & 0.22 & \nodata & 3.82 & 0.50 & \nodata & 24.7 &
7.1 & \nodata & 40 & 12 & \nodata \nl
~~~~~~MMS2 & 2.55 & 0.21 & \nodata & 3.42 & 0.45 & \nodata & 20.1 &
5.7 & \nodata & 29.4 & 8.9 & \nodata \nl
~~~~~~MMS3 & 1.08 & 0.10 & \nodata & 1.74 & 0.23 & \nodata & 9.2 &
2.4 & \nodata & 13.3 & 4.1 & \nodata \nl
HH211-mm & 5.73 & 0.47 & 21 & 8.2 & 1.1 & 33 & 34.7 & 7.0 & 110 & 51
& 16 & 390 \\
L1551~IRS5 & 12.1 & 1.0 & 22 & 18.2 & 2.4 & 37 & 94 & 19 & 350 & 164
& 49 & 630 \\
HL~Tau & 2.97 & 0.25 & 16 & 4.04 & 0.54 & 20 & 16.8 & 4.8 & 150 &
26.4 & 8.7 & 400 \\
TMC1A & 1.80 & 0.15 & \phn8 & 2.17 & 0.30 & 17 & 12.7 & 2.6 & \phn92
& 20.6 & 6.6 & 300 \nl
L1527 & 5.90 & 0.48 & 15 & 8.4 & 1.1 & 21 & 44.8 & 9.0 & 130 & 66
& 20 & 340 \\
\enddata
\tablenotetext{a}{For most sources the flux density is integrated
over an area 45~arcsec in radius, centered on the central
protostar.  For L1448-mm a radius of 40~arcsec is used to avoid
including emission from L1448N to the north.  For L1448N we used
18~arcsec, and for L1448NW the flux is the peak flux per beam
obtained from images smoothed to the resolution of the 850~$\mu$m
data using the model beams listed in Table~\ref{beam-fits}.  For
NGC1333~IRAS2 we also quote integrated flux densities for IRAS2:CR1
and IRAS2:CR2 separately, using radii of 20~arcsec and 16.5~arcsec
respectively. See sections~\ref{spectra} and \ref{svs13} for a
description of the sources MMS1, MMS2, and MMS3.}
\tablenotetext{b}{1$\sigma$ rms noise for the images shown in
Fig.~\ref{scuba-images}, in mJy~beam$^{-1}$.}
\end{deluxetable}

$$\psfig{file=fig01.ps,height=8in,angle=0}$$

\vspace{-8in}
\begin{sideways}
\parbox{8in}{\vspace{4.5in}\centerline{Fig.~1.}}
\end{sideways}

$$\psfig{file=fig02_1.ps,height=8in,angle=0}$$

\vspace{-8in}
\begin{sideways}
\parbox{8in}{\vspace{5.5in}\centerline{Fig.~2.}}
\end{sideways}

$$\psfig{file=fig02_2.ps,height=8in,angle=0}$$

\vspace{-8in}
\begin{sideways}
\parbox{8in}{\vspace{5.5in}\centerline{Fig.~2 (cont).}}
\end{sideways}

$$\psfig{file=fig02_3.ps,height=8in,angle=0}$$

\vspace{-8in}
\begin{sideways}
\parbox{8in}{\vspace{5.5in}\centerline{Fig.~2 (cont).}}
\end{sideways}

$$\psfig{file=fig02_4.ps,height=8in,angle=0}$$

\vspace{-8in}
\begin{sideways}
\parbox{8in}{\vspace{5.5in}\centerline{Fig.~2 (cont).}}
\end{sideways}

$$\psfig{file=fig03.ps,height=8in,angle=0}$$

\vspace{-8in}
\begin{sideways}
\parbox{8in}{\vspace{5.5in}\centerline{Fig.~3.}}
\end{sideways}

$$\psfig{file=fig04_1.ps,height=7.5in,angle=0}$$

\centerline{Fig.~4.}

$$\psfig{file=fig04_2.ps,height=7.5in,angle=0}$$

\centerline{Fig.~4 (cont).}

$$\psfig{file=fig04_3.ps,height=7.5in,angle=0}$$

\centerline{Fig.~4 (cont).}

$$\psfig{file=fig04_4.ps,height=3.75in,angle=0}$$

\centerline{Fig.~4 (cont).}

$$\psfig{file=fig05.ps,height=7.5in,angle=0}$$

\centerline{Fig.~5.}

$$\psfig{file=fig06.ps,height=3.75in,angle=0}$$

\centerline{Fig.~6.}

$$\psfig{file=fig07.ps,height=7.5in,angle=0}$$

\centerline{Fig.~7.}

$$\psfig{file=fig08.ps,height=7.5in,angle=0}$$

\centerline{Fig.~8.}

\hspace{-0.5in}
\psfig{file=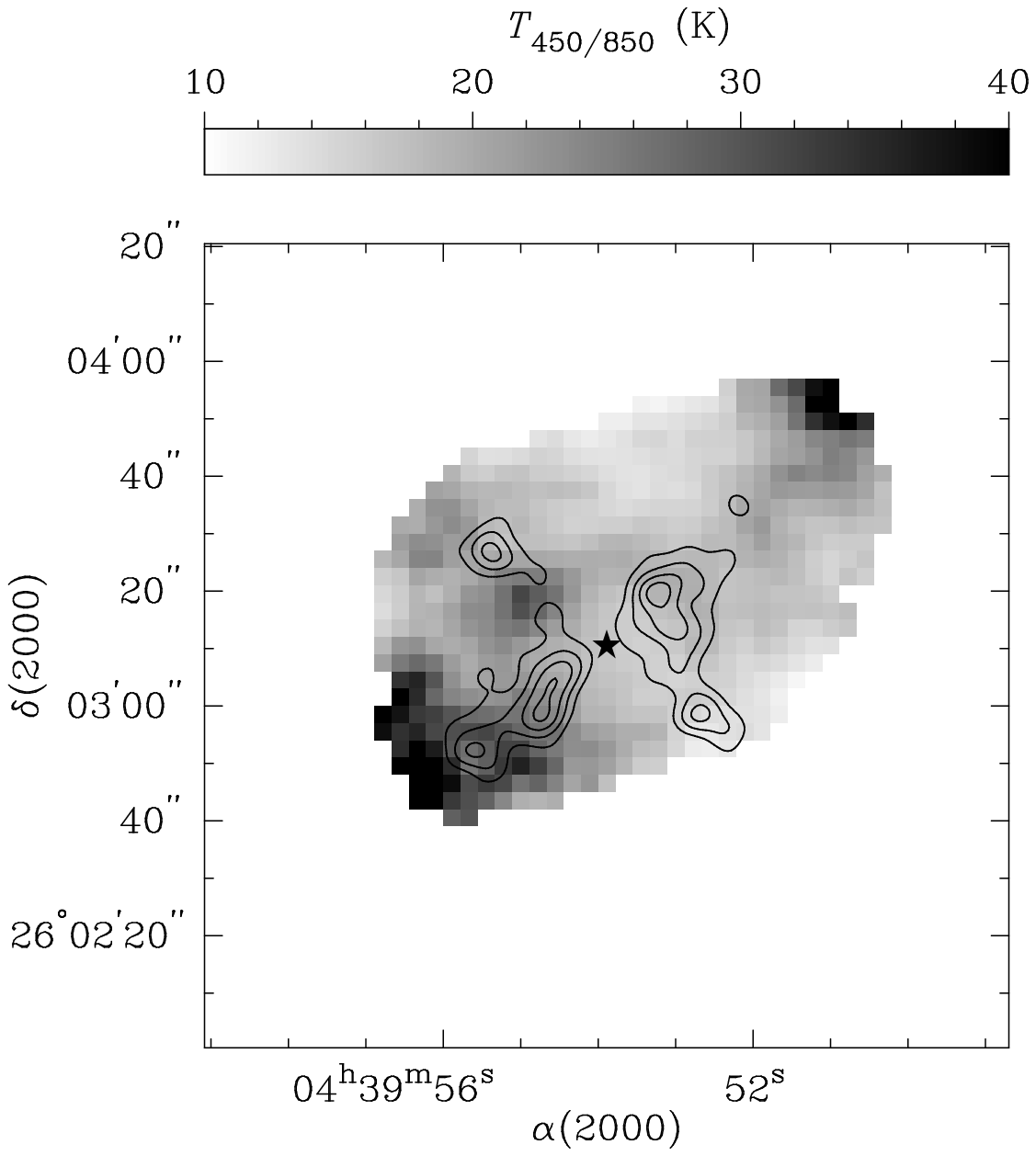,height=8in,angle=0}

\vspace{-2.8in}
\centerline{Fig.~9.}

\newpage

$$\psfig{file=fig10.ps,height=3.5in,angle=-90}$$

\centerline{Fig.~10.}

\end{document}